\documentclass{article}
\usepackage{spconf}
\usepackage{blindtext, graphicx}
\usepackage[cmex10]{amsmath}
\usepackage{bm}
\usepackage{amssymb,amsthm,mathrsfs,amsfonts,amsmath}
\usepackage{bbm}
\usepackage{caption}

\PassOptionsToPackage{obeyspaces}{url}
\PassOptionsToPackage{bookmarks=false}{hyperref}

\usepackage[nospace, nocompress]{cite}

\makeatletter
\def\hlinewd#1{\noalign{\ifnum0=`}\fi\hrule \@height #1 \futurelet\reserved@a\@xhline}
\makeatother

\title{Bayesian Non-Parametric Multi-Source Modelling based\\ Determined Blind Source Separation}
\vspace{-15pt}
\name{Chaitanya Narisetty, Tatsuya Komatsu, \textnormal{and} Reishi Kondo}
\address{Data Science Research Laboratories, NEC Corporation, Japan}
\begin{document}
\maketitle

\begin{abstract}
This paper proposes a determined blind source separation method using Bayesian non-parametric modelling of sources.
Conventionally source signals are separated from a given set of mixture signals by modelling them using non-negative matrix factorization (NMF).
However in NMF, a latent variable signifying model complexity must be appropriately specified to avoid over-fitting or under-fitting.
As real-world sources can be of varying and unknown complexities, we propose a Bayesian non-parametric framework which is invariant to such latent variables.
We show that our proposed method adapts to different source complexities, while conventional methods require parameter tuning for optimal separation. 
\end{abstract}

\vspace{-3pt}
\section{Introduction}\vspace{-1pt}
\label{sec:Intro}
The ubiquity of smart devices in our present day provide a diverse range of audio based applications like detecting threat indicating sounds, transcribing speech, building AI assistants etc. to facilitate our day-to-day activities \cite{foote1999overview}. Depending on the application, our desired information extracted from the audio signals changes. Real world audio signals are often obscured by undesired information and necessitate the development of source separation methods \cite{davies2002audio}. As most smart devices are equipped with two or more microphones, the recorded multi-channel data can be leveraged to improve the separation performance \cite{weinstein1993multi}. However the arrangement of microphones varies across devices, as do the source characteristics and locations.  Separating sources without such device or source information is termed as blind source separation (BSS) \cite{comon2010handbook, cao1996general}.\vspace{5pt}

Fundamental BSS techniques include matrix factorization methods like independent component analysis (ICA) \cite{comon1994independent}, non-negative matrix factorization (NMF) \cite{ozerov2011multichannel} etc. These techniques formulate the audio signals captured by microphones as linear mixtures of two or more source signals. In other words, the spectra of mixture signals are treated as linear combinations of spectra of source signals, and solved by assuming independence among the distributions of source spectra. This assumption of independence is often not true because the spectra of a typical audio source is highly correlated with itself across different time intervals. Such correlations or fundamental patterns in each source are efficiently extracted using NMF \cite{nari2018Modelling}, which are referred to as bases. The way these bases are linearly combined to reconstruct spectra, are referred to as activations and  are extracted simultaneously by NMF. BSS methods are therefore extended by modelling such correlations using multi-channel NMF \cite{ozerov2011multichannel, sawada2013multichannel}. It is intuitive that source separation becomes more reliable as the number of available microphones increases. When there are as many number of microphones as there are sources, the situation is said to be determined. Ono et. al. \cite{ono2011stable} proposed independent vector analysis (IVA) for determined BSS using an auxiliary function technique to derive stable and fast update rules for demixing parameters. Kitamura et. al. \cite{kitamura2016determined} proposed an independent low-rank matrix analysis method (ILRMA) by unifying IVA and multi-source NMF modelling.\vspace{2pt}

In the above discussed NMF-based BSS methods, it is required to provide an appropriate value for NMF's model complexity i.e. the number of bases extracted for each source model. This latent variable when too low or too high, leads to under-fitting or over-fitting respectively. Its value is often chosen depending on source characteristics or by parameter tuning. However real world source characteristics are unknown, and their complexities range from simple keyboard clicking to moderate sounds of drums to complex music pieces. Although non-parametric methods exist for estimating number of sources, to the best of our knowledge, there aren't any NMF-based BSS techniques which can adaptively model sources having different complexities.\vspace{2pt}

We propose a non-parametric framework of multi-source modelling unified with IVA for determined BSS, and overcome the problem of tuning NMF's model complexity parameter. Our proposed method utilizes the concepts of variational Bayesian inference to statistically estimate each source's complexity, thereby optimally separating the sources. Proposed method therefore serves as a generalization of ILRMA by adapting to varying source complexities.\vspace{2pt}

The remainder of this paper is organized as follows. Section~\ref{sec:ConvMethod} details the conventional method: ILRMA. Section~\ref{sec:PropMeth} formulates our proposed method. Section~\ref{sec:SimsResults} details the experimental results and discusses a few potential implications of the proposed method. Finally section~\ref{sec:Conc} concludes our work.
\section{Conventional Method}
\label{sec:ConvMethod}
Most BSS methods model the given mixture spectra as linear combinations of source spectra and are formulated as
\begin{align}
\label{eq:mixing}
\pmb{x}_{ij} = \pmb{A}_i \pmb{s}_{ij},
\end{align}
where $\pmb{x}_{ij}=(x_{ij,1},\dots, x_{ij, M})^\mathsf{T}$ and $\pmb{s}_{ij} = (s_{ij,1},\dots, s_{ij,N})^\mathsf{T}$ denote the mixture and source spectra respectively for each index $i\in \{1,2,\dots,I\}$, $j\in \{1,2,\dots,J\}$. $(.)^\mathsf{T}$ denotes a transpose and $I,J,M,N$ denote the number of frequency bins, time frames, microphones and sources respectively. $\textbf{\textit{A}}_i$ is an $M\times N$ mixing matrix comprising of $N$ steering vectors for the $N$ respective sources. In a determined case, $M=N$ and the square matrix $\pmb{A}_i$ has a valid inverse matrix. Therefore \cite{ono2011stable} proposed IVA by defining a demixing matrix $\pmb{W}_i = \pmb{A}_i^{-1} = (\pmb{w}_{i,1},\dots,\pmb{w}_{i,M})^\mathsf{H}$ and reformulated Eq.~(\ref{eq:mixing}) as
\begin{align}
\label{eq:demixing}
\pmb{y}_{ij} = \pmb{W}_i \pmb{x}_{ij},
\end{align}
where $\pmb{y}_{ij}=(y_{ij,1},\dots, y_{ij, M})^\mathsf{T}$ denotes the estimated source spectra and $(.)^\mathsf{H}$ denotes a hermitian transpose.
\vspace{-5pt}

\subsection{ILRMA: Unifying IVA and NMF}
This method extends IVA by independently modelling each source with an isotropic complex Gaussian distribution. For each source index $m \in \{1,\dots,M\}$, the distribution variance denoted as $r_{ij,m}$ is non-negative and modelled using NMF as\vspace{-2pt}
\begin{align}
\label{eq:sourceModel}
r_{ij,m} = \sum_{k=1}^{K_m} t_{ik,m}v_{kj,m},
\end{align}
where $t_{ik,m}$ and $v_{kj,m}$ are the elements of basis and activation vectors respectively. $K_m$ is NMF's model complexity parameter, signifying the number of basis vectors for $m^{th}$ source. The cost function $Q$ of ILRMA is derived in \cite{kitamura2016determined} as \vspace{-2pt}
\begin{align}
\label{eq:costFuncILRMA}
Q=-2J\sum_i|\det \pmb{W}_i| + \sum_{i,j,m}\bigg[ \log r_{ij,m}+\frac{|y_{ij,m}|^2}{r_{ij,m}} \bigg].
\end{align}
ILRMA estimates the demixing parameters by maximizing cost function $Q$ using multi-source NMF modelling in Eq.~(\ref{eq:sourceModel}).
\vspace{-17pt}

\subsection{Limitation: Number of Source Bases}
In the formulation of ILRMA, note that a set of complexity parameters $\{K_m, 1\leq m\leq M\}$ is to be specified by the user. Failing to provide a reasonable estimate of each source's complexity will lead to over-fitting or under-fitting. This is discussed in \cite{kitamura2016determined} by comparing the cases of separating speech signals with that of music signals. The former requires only $2$ bases for optimal modelling as compared to the latter requiring more than $30$. Therefore the model complexity parameter can effect separation performance, depending on the types of sources being separated. For simplicity, ILRMA assumes equal number of bases for each source i.e. $\{K_1 = \dots = K_M = K\}$. This further limits ILRMA's ability to optimally separate a low complexity source and a high complexity source from their mixture signals.
\vspace{-7pt}
\section{Proposed Multi-Source Modelling}\vspace{-3pt}
\label{sec:PropMeth}
We overcome the limitation of ILRMA by proposing a probabilistic modelling of the variance of source distributions. In such techniques, it is common to introduce hidden variables to capture the structure of given observed data, and utilize inference algorithms to estimate the posterior distribution. Accordingly, our proposed method flexibly models each source using a large number of basis vectors $K$ and incorporates a \textit{reliability} value for each basis vector. We denote each reliability value as $z_{k,m}$ which can be interpreted as a quantified contribution of $k^{th}$ basis towards $m^{th}$ source. 
\vspace{-10pt}
\setlength{\abovedisplayskip}{4pt}
\setlength{\belowdisplayskip}{5pt}
\subsection{Model Formulation}\vspace{-2pt}
Contrast to the NMF-based source modelling in Eq.~(\ref{eq:sourceModel}), we propose a probabilistic model for source variance $r_{ij,m}$ as
\begin{align}
\label{eq:propSourceModel}
r_{ij,m} = \sum_{k=1}^{K} z_{k,m}t_{ik,m}v_{kj,m}.
\end{align}
where the prior distributions for each of $t_{ik,m}$, $v_{kj,m}$ and $z_{k,m}$ are drawn from a random process as
\begin{align}
\label{eq:propPrior}
p(t_{ik,m}) \;&\sim\; \text{Gamma}(a_0, a_0),\nonumber\\
p(v_{kj,m}) \;&\sim\; \text{Gamma}(b_0, b_0),\nonumber\\
p(z_{k,m}) \;&\sim\; \text{Gamma}(c_0, c_m),
\vspace{-5pt}\end{align}
where $a_0, b_0, c_0$ are positive constants, Gamma$(.,.)$ is a gamma distribution defined over a shape parameter and a rate (inverse-scale) parameter. 
When $c_0 \ll 1$, a sparse prior is set over the reliability values and therefore adapts to different model complexities depending on each source's characteristics \cite{tan2009automatic}. As each source's expected variance should correspond to the expectation of its power, the choice of prior parameters require that $\mathbb{E}_p[|y_{ij,m}|^2] = \mathbb{E}_p[r_{ij,m}]$,
\begin{align}
\label{eq:cm}
\Rightarrow \mathbb{E}_p[|y_{ij,m}|^2] &= \sum_k \mathbb{E}_p[z_{k,m}] \mathbb{E}_p[t_{ik,m}v_{kj,m}] = \sum_k (c_0/c_m), \nonumber\\
\Rightarrow \quad \quad \quad \;\, \, c_m &= c_0 K \big[\sum_i\sum_j |y_{ij,m}|^2/(IJ)\big]^{-1}.
\end{align}
Maximizing the cost function in Eq.~(\ref{eq:costFuncILRMA}), as updated by Eq.~(\ref{eq:propSourceModel}) is the central focus of our approach.
\vspace{-9pt}
\subsection{Variational Bayesian Inference}
The isotropic gaussian distribution of sources is not conjugate to the gamma distribution of source parameters $t, v, z$. This non-conjugacy precludes the use of Markov chain Monte Carlo (MCMC) based Metropolis-Hastings, Gibbs sampling for inferring our desired posterior \cite{hastings1970monte, blei2007correlated}. However variational approaches have provided alternatives for non-conjugate models that are also faster for large amounts of data. We adopt a fully factorized mean-field variational inference technique as it assumes conditional independence among the hidden variables \cite{hoffman2013stochastic} and approximates them from a family of conditional distributions over variational parameters \cite{jordan1999introduction}. Generalized inverse Gaussian (GIG) distributions are chosen as our variational family and are expressed as \vspace{3pt}
\setlength{\belowdisplayskip}{7pt}
\begin{align}
\label{eq:GIG}
\text{GIG}(\theta|\gamma, \rho, \tau) = \frac{\exp\{(\gamma-1)\log\theta-\rho\theta-\tau/\theta\}}{2(\tau/\rho)^{\gamma/2}\mathcal{K}_\gamma(2\sqrt{\rho\tau})},
\end{align}
where $\mathcal{K}(.)$ is a modified Bessel function of the second kind and $\gamma, \rho, \tau$ are the variational hyper-parameters. GIG's sufficient statistics includes $(1/\theta)$ which eases the optimization of our cost function's $(1/r_{ij,m})$ term and motivates our choice of GIG \cite{blei2010bayesian}. We define the conditional distributions as
\begin{align}
\label{eq:propCondDist}
q(t_{ik,m}|\Theta_{\setminus t_{ik,m}}) \;&\sim\; \text{GIG}(a_0, \rho_{ik,m}^{(t)}, \tau_{ik,m}^{(t)}) ,\nonumber\\
q(v_{kj,m}|\Theta_{\setminus v_{kj,m}}) \;&\sim\; \text{GIG}(b_0, \rho_{kj,m}^{(v)}, \tau_{kj,m}^{(v)}) ,\nonumber\\
q(z_{k,m}|\Theta_{\setminus z_{k,m}}) \;&\sim\; \text{GIG}(c_0, \rho_{k,m}^{(z)}, \tau_{k,m}^{(z)}).
\end{align}
We now derive update equations from the cost function in Eq.~(\ref{eq:costFuncILRMA}) using first-order Taylor expansion and Jensen's inequality \cite{lafferty2006correlated} by introducing their respective auxiliary positive constants $\alpha_{ij,m}$ and $\beta_{ijk,m}$ as\setlength{\belowdisplayskip}{10pt}
\begin{align}
\label{eq:costDerivation}
Q &+ 2J\sum_i|\det \pmb{W}_i|= \sum_{i,j,m}\bigg[\log r_{ij,m} + \frac{|y_{ij,m}|^2}{r_{ij,m}}\bigg],\quad\nonumber\\
&\leq \sum_{i,j,m}\mathbb{E}_q\bigg[\log r_{ij,m} + \frac{|y_{ij,m}|^2}{r_{ij,m}}\bigg] + \mathbb{E}_q\bigg[\log \frac{q(t|\Theta_{\setminus t})}{p(t|a_0)}\bigg]\nonumber\\
&\quad +\mathbb{E}_q\bigg[\log \frac{q(v|\Theta_{\setminus v})}{p(v|b_0)}\bigg] + \mathbb{E}_q\bigg[\log \frac{q(z|\Theta_{\setminus z})}{p(z|c_0,c_m)}\bigg], \nonumber\\
&\leq \sum_{i,j}\bigg[\sum_{k,m} |y_{ij,m}|^2 \beta_{ijk,m}^2 \mathbb{E}_q\big[z_{k,m}^{-1}t_{ik,m}^{-1}v_{kj,m}^{-1}\big]\nonumber\\
&\quad - 1+\log\alpha_{ij,m} + \frac{1}{\alpha_{ij,m}}\sum_k \mathbb{E}_q[z_{k,m}t_{ik,m}v_{kj,m}] \bigg] \nonumber\\ 
&\quad + \mathbb{E}_q[- \rho_{ik,m}^{(t)}t_{ik,m} - \tau_{ik,m}^{(t)}/t_{ik,m} + a_0 t_{ik,m}]\nonumber\\
&\quad + \mathbb{E}_q[- \rho_{kj,m}^{(v)}v_{kj,m} - \tau_{kj,m}^{(v)}/v_{kj,m} + b_0 v_{kj,m}]\nonumber\\
&\quad + \mathbb{E}_q[- \rho_{k,m}^{(z)}z_{k,m} - \tau_{k,m}^{(z)}/z_{k,m} + c_m z_{k,m}]+C,
\end{align}
where $C$ denotes a leftover constant. Note that the pairs $(t,t^{-1}), (v,v^{-1})$ and $(z,z^{-1})$ are sufficient statistics for their respective GIG distributions. This allows us to avoid taking partial derivatives and directly derive the analytic coordinate ascent updates for our hyper-parameters by comparing the coefficients of sufficient statistics \cite{blei2010bayesian}. Constants $\alpha_{ij,m}$ and $\beta_{ijk,m}$ re-tighten the above inequality~(\ref{eq:costDerivation}) when:
\begin{align}
\label{eq:alpha}
\alpha_{ij,m} &= \sum_k \mathbb{E}_q[z_{k,m}] \mathbb{E}_q[t_{ik,m}] \mathbb{E}_q[v_{kj,m}],\\
\label{eq:beta}
\beta_{ijk,m} &= \frac{\mathbb{E}_q\big[z_{k,m}^{-1}\big] \mathbb{E}_q\big[t_{ik,m}^{-1}\big] \mathbb{E}_q\big[v_{kj,m}^{-1}\big]}{\sum_k \mathbb{E}_q\big[z_{k,m}^{-1}\big] \mathbb{E}_q\big[t_{ik,m}^{-1}\big] \mathbb{E}_q\big[v_{kj,m}^{-1}\big]}.
\end{align}
Expectation of source parameters in Eqs.~(\ref{eq:alpha}) and~(\ref{eq:beta}) can be obtained similar to the expectation of a random variable $\theta$ defined over a GIG distribution in Eq.~(\ref{eq:GIG}) as
\begin{align}
\label{eq:meanBessel}
\mathbb{E}[\theta]=\frac{\mathcal{K}_{\gamma+1}(2\sqrt{\rho\tau})\sqrt{\tau}}{\mathcal{K}_\gamma(2\sqrt{\rho\tau})\sqrt{\rho}}, \mathbb{E}\bigg[\frac{1}{\theta}\bigg]=\frac{\mathcal{K}_{\gamma-1}(2\sqrt{\rho\tau})\sqrt{\rho}}{\mathcal{K}_\gamma(2\sqrt{\rho\tau})\sqrt{\tau}}.
\end{align}
\setlength{\belowdisplayskip}{5pt}
The update equations for our hyper-parameters are derived as\setlength{\abovedisplayskip}{7pt}
\begin{align}
\label{eq:updateRules_rhotau_1}
\rho_{ik,m}^{(t)} &= a_0+\mathbb{E}_q[z_{k,m}] \sum_j \mathbb{E}_q[v_{kj,m}]\alpha_{ij,m}^{-1},\\
\tau_{ik,m}^{(t)} &= \sum_j |y_{ij,m}|^2 \beta_{ijk,m}^2\mathbb{E}_q\big[z_{k,m}^{-1}\big] \mathbb{E}_q\big[v_{kj,m}^{-1}\big],\\
\rho_{kj,m}^{(v)} &= b_0+\mathbb{E}_q[z_{k,m}] \sum_i \mathbb{E}_q[t_{ik,m}]\alpha_{ij,m}^{-1},\\
\tau_{kj,m}^{(v)} &= \sum_i |y_{ij,m}|^2 \beta_{ijk,m}^2\mathbb{E}_q\big[z_{k,m}^{-1}\big] \mathbb{E}_q\big[t_{ik,m}^{-1}\big],\\
\rho_{k,m}^{(z)}\; &= c_m+\sum_i\sum_j \mathbb{E}_q[t_{ik,m}] \mathbb{E}_q[v_{kj,m}]\alpha_{ij,m}^{-1},\\
\label{eq:updateRules_rhotau_2}
\tau_{k,m}^{(z)}\; &= \sum_i\sum_j|y_{ij,m}|^2 \beta_{ijk,m}^2 \mathbb{E}_q\big[t_{ik,m}^{-1}\big] \mathbb{E}_q\big[v_{kj,m}^{-1}\big].
\end{align}
\vspace{-17pt}
\subsection{Update Rules for Demixing Matrix}\vspace{-2pt}
Given each source's variance, the proposed method does not alter the partial derivatives of cost function $Q$ over the demixing matrix $\pmb W_i$. Hence its update equations coincide with those described for IVA using an auxiliary function technique \cite{ono2011stable} and are derived as follows:
\begin{align}
\label{eq:updateRules_W_1}
V_{i,m} &= \frac{1}{J}\sum_j\frac{1}{r_{ij,m}}\pmb{x}_{ij}\pmb{x}_{ij}^h, \\
\label{eq:updateRules_W_2}
\pmb{w}_{i,m} &\leftarrow (\pmb{W}_iV_{i,m})^{-1}\pmb{e}_m, \\
\label{eq:updateRules_W_3}
\pmb{w}_{i,m} &\leftarrow \pmb{w}_{i,m}(\pmb{w}_{i,m}^hV_{i,m}\pmb{w}_{i,m})^{-1/2},
\end{align}
where $\pmb{e}_m$ is a unit vector whose $m^{th}$ element equals one. After the elements of demixing matrix are estimated, the separated source spectra can be extracted as\setlength{\belowdisplayskip}{7pt}
\begin{align}
\label{eq:sepSources}
y_{ij,m} \leftarrow \pmb{w}_{i,m}^h\pmb{x}_{ij}.
\end{align}

In each iteration of the proposed method, hyper-parameters $\rho^{(.)}$ and $\tau^{(.)}$ are updated using Eq.~(\ref{eq:updateRules_rhotau_1})-(\ref{eq:updateRules_rhotau_2}). Demixing matrices $\pmb{W}_i$ and the separated source spectra are then updated using Eq.~(\ref{eq:updateRules_W_1})-(\ref{eq:sepSources}). As each source's modelling begins with a large $K$, our variational inference is computationally exacting. However over a few iterations, the sparse prior placed over $z$ estimates a small number of bases which are reliable. So we employ a thresholding technique \cite{paisley2009nonparametric} to skip the optimization of less reliable bases in subsequent iterations.
\vspace{-5pt}
\section{Simulations and Results}
\label{sec:SimsResults}\vspace{-4pt}
\subsection{Experimental Conditions}
\begin{figure}[t]
	\centering
	\includegraphics[clip, trim=6cm 7.5cm 6cm 3cm, width=0.8\linewidth]{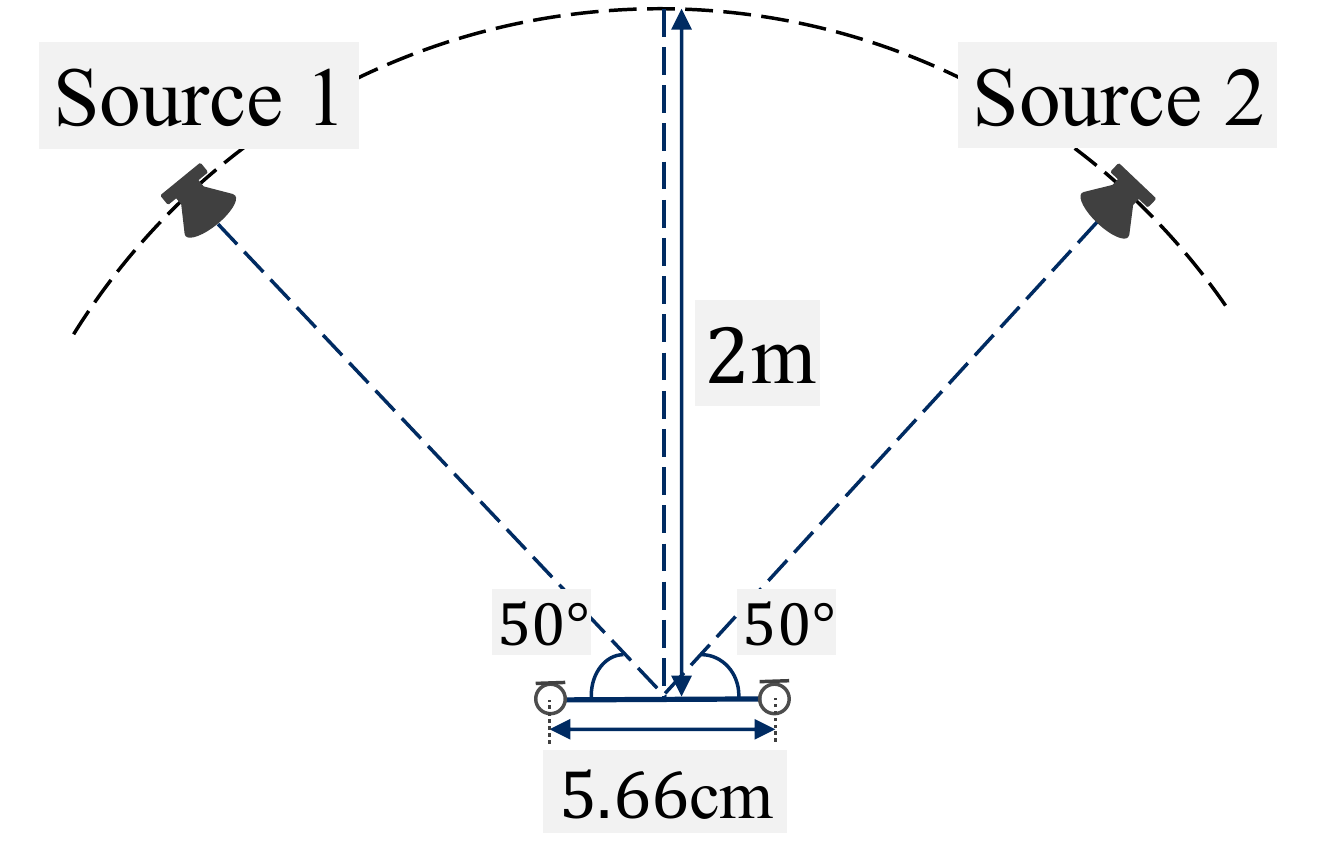}
	\caption{E2A recording conditions (reverberation time: $300$ms)}
	\label{fig:RIR}
	\vspace{-10pt}
\end{figure}

\vspace{-2pt}
We evaluate our proposed method on the DSD$100$ dataset containing $100$ professionally produced songs from the 2016 SiSEC challenge \cite{SiSEC17}. Each song data consists of clean sounds for vocals, drums, bass and other accompaniments, and lasts for more than $2$-$3$ minutes. So we only choose a $30$ second portion ($30$s-$60$s time interval) of clean sources and down-sample them to $16\,$kHz for creating mixture signals. For each of these $100$ songs, we randomly choose between the pairs (drums, vocals) or (bass, vocals) and create synthetic two-channel reverberant mixture signals using the recoding conditions shown in Fig.~\ref{fig:RIR}. Room impulse responses E2A  ($T_{60}=300\,$ms) for above recording conditions were obtained from the RWCP Sound Scene Database \cite{nakamura2000acoustical}.\vspace{5pt}

Mixture spectra are estimated from the time domain signals using a Hamming window of length $512\,$ms shifted every $128\,$ms, which were found to be optimal by \cite{kitamura2017experimental}. Each demixing matrix $\pmb W_i$ is initialized with an identity matrix. We model each source's variance with $K=30$ bases and set $a_0=b_0=0.1, c_0=1/K$. Hyper-parameters $\rho^{(.)}$ and $\tau^{(.)}$ are initialized randomly from gamma distributions with shape and rate parameters set to $1000$. Parameters are optimized for $100$ iterations and then the separated spectra are converted to time domain using a back-projection technique \cite{murata2001approach}.
\vspace{-12pt}
\subsection{Evaluations and Comparisons}\vspace{-2pt}
Three metrics: signal to distortion ratio (SDR), signal to interference ratio (SIR) and signal to artifacts ratio (SAR) \cite{vincent2006performance} are used to evaluate the quality of separated sources. Separation performance of the proposed method is compared with three NMF-based BSS method i.e. IVA \cite{ono2011stable}, MNMF \cite{sawada2013multichannel} and ILRMA \cite{kitamura2016determined} (with $5$ source bases). Each separation is repeated for $10$ different random initializations, and the average of the above performance metrics are reported in Table~\ref{tab:SDIAR}.\vspace{-3pt}
\begin{table}[th]
	\centering
	\caption{Comparison of Separation Performance\vspace{-5pt}}
	\label{tab:SDIAR}
	\bgroup\normalsize
	\def\arraystretch{1.3}%
	\begin{tabular*}{0.9\linewidth}{ l | c c c}
		\hlinewd{1.25pt}
		Methods & \hspace{12pt}SDR\hspace{10pt} & \hspace{10pt}SIR\hspace{10pt} & \hspace{13pt}SAR\hspace{13pt} \\
		\hline
		IVA \cite{ono2011stable} & $2.6\,$dB & $7.4\,$dB & $5.7\,$dB \\
		MNMF\cite{sawada2013multichannel} & $3.5\,$dB & $8.6\,$dB & $6.2\,$dB \\
		ILRMA \cite{kitamura2016determined}\hspace{4pt} & $8.7\,$dB & $16.2\,$dB & $12.3\,$dB \\
		\textbf{Proposed} & $\pmb{9.9}\,$dB & $\pmb{17.4}\,$dB & $\pmb{13.2}\,$dB \\
		\hlinewd{1.2pt}
	\end{tabular*}
	\egroup
	\vspace{-0pt}
\end{table}

\begin{figure}[th]
	\centering
	\includegraphics[clip, trim=0cm 0cm 0cm 0cm, width=\linewidth]{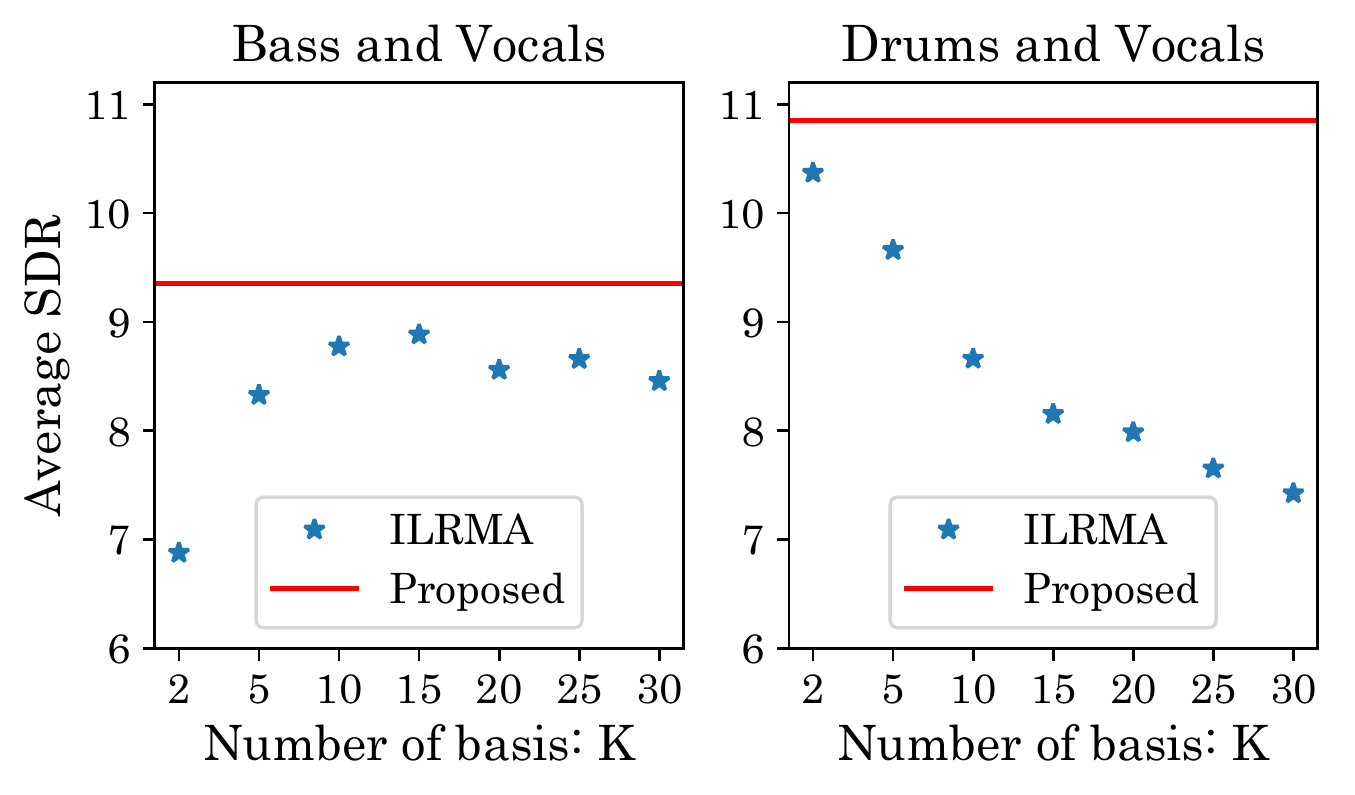}
	\caption{SDR for ILRMA using different number of bases as compared with proposed method's SDR for mixture signals of bass and vocals (left), and of drums and vocals (right).}
	\label{fig:CompPropILRMA}
	\vspace{-10pt}
\end{figure}

Although the proposed method outperforms ILRMA, it is important to verify that we overcome its limitation of having to tune the complexity parameter $K$. This limitation can be seen in Fig.~\ref{fig:CompPropILRMA} where ILRMA's SDR averaged over all the mixture signals containing $\,$(bass, vocals) increases as $K$ increases, while an opposite trend is seen for mixture signals containing (drums, vocals). The proposed method on the other hand starts with a large value for $K$ ($=30$ in this case) and tunes itself depending on each source's characteristics. Hence it optimally separates the sources from both types of mixture signals. The choice of $K$, if sufficiently large, does not to significantly impact proposed method's performance.\vspace{-5pt}

\vspace{-10pt}
\subsection{Possible Extension}\vspace{-2pt}
As each source is modelled individually using NMF, it is possible to extend our approach by considering a common basis and activation matrix for capturing the variance of all sources i.e. $r_{ij,m}=\sum_k z_{k,m} t_{ik} v_{kj}$. This requires fewer parameters to be estimated and its computational complexity becomes at least half as compared to that of the proposed method. We note that this extension is similar to Ozerov's MNMF \cite{ozerov2011multichannel} and has also been considered for ILRMA \cite{kitamura2016determined}. However the key difference is that we do not restrict the overall contribution of each basis to be $1$, i.e. $\sum_m z_{k,m} = 1 \,\forall\,k$. Due to space constraints, this extension will be explored as a future work.

\section{Conclusions}
\label{sec:Conc}
This work proposes a Bayesian generalization of ILRMA for determined blind source separation by performing multi-source modelling using non-parametric NMF. Our formulation for individual source modelling is able to overcome the limitation of conventional method, whose separation performance is effected by NMF's model complexity parameter. Proposed approach is flexible in modelling sources of different complexities and is therefore able to optimally separate them. We further show that our approach outperforms the state-of-the-art NMF based techniques.

\bibliographystyle{IEEEbib}
\bibliography{mybib}

\begin{thebibliography}{10}

\bibitem{foote1999overview}
J.~Foote,
\newblock ``An overview of audio information retrieval,''
\newblock {\em Multimedia systems}, vol.7, no.1, pp.2--10, 1999.

\bibitem{davies2002audio}
M.~Davies,
\newblock ``Audio source separation,''
\newblock in {\em Institute of mathematics and its applications conference
  series}, 2002, vol.~71, pp. 57--68.

\bibitem{weinstein1993multi}
E.~Weinstein, M.~Feder, and A.~V. Oppenheim,
\newblock ``{Multi-channel} signal separation by decorrelation,''
\newblock {\em IEEE transactions on Speech and Audio Processing}, vol.1, no.4,
  pp.405--413, 1993.

\bibitem{comon2010handbook}
P.~Comon and C.~Jutten,
\newblock {\em Handbook of Blind Source Separation: Independent component
  analysis and applications},
\newblock Academic press, 2010.

\bibitem{cao1996general}
X.~Cao and R.~Liu,
\newblock ``General approach to blind source separation,''
\newblock {\em IEEE Transactions on signal Processing}, vol.44, no.3,
  pp.562--571, 1996.

\bibitem{comon1994independent}
P.~Comon,
\newblock ``Independent component analysis, a new concept?,''
\newblock {\em Signal processing}, vol.36, no.3, pp.287--314, 1994.

\bibitem{ozerov2011multichannel}
A.~Ozerov, C.~F{\'e}votte, R.~Blouet, and J.~Durrieu,
\newblock ``Multichannel nonnegative tensor factorization with structured
  constraints for {user-guided} audio source separation,''
\newblock in {\em IEEE International Conference on Acoustics, Speech and Signal
  Processing (ICASSP)}, 2011, pp. 257--260.

\bibitem{nari2018Modelling}
C.~Narisetty, T.~Komatsu, and R.~Kondo,
\newblock ``Modelling of sound events with hidden imbalances based on
  clustering and separate {Sub-Dictionary} learning,''
\newblock in {\em 26th European Signal Processing Conference (EUSIPCO)}, 2018.

\bibitem{sawada2013multichannel}
H.~Sawada, H.~Kameoka, S.~Araki, and N.~Ueda,
\newblock ``Multichannel extensions of {non-negative} matrix factorization with
  {complex-valued} data,''
\newblock {\em IEEE Transactions on Audio, Speech, and Language Processing},
  vol.21, no.5, pp.971--982, 2013.

\bibitem{ono2011stable}
N.~Ono,
\newblock ``Stable and fast update rules for independent vector analysis based
  on auxiliary function technique,''
\newblock in {\em IEEE Workshop on Applications of Signal Processing to Audio
  and Acoustics (WASPAA)}, 2011, pp. 189--192.

\bibitem{kitamura2016determined}
D.~Kitamura, N.~Ono, H.~Sawada, H.~Kameoka, and H.~Saruwatari,
\newblock ``Determined blind source separation unifying independent vector
  analysis and nonnegative matrix factorization,''
\newblock {\em IEEE/ACM Transactions on Audio, Speech and Language Processing
  (TASLP)}, vol.24, no.9, pp.1622--1637, 2016.

\bibitem{tan2009automatic}
V.~Y. Tan and C.~F{\'e}votte,
\newblock ``Automatic relevance determination in nonnegative matrix
  factorization,''
\newblock in {\em SPARS'09-Signal Processing with Adaptive Sparse Structured
  Representations}, 2009.

\bibitem{hastings1970monte}
W.~K. Hastings,
\newblock ``Monte carlo sampling methods using markov chains and their
  applications,''
\newblock 1970.

\bibitem{blei2007correlated}
D.~M. Blei, J.~D. Lafferty, et~al.,
\newblock ``A correlated topic model of science,''
\newblock {\em The Annals of Applied Statistics}, vol.1, no.1, pp.17--35, 2007.

\bibitem{hoffman2013stochastic}
M.~D. Hoffman, D.~M. Blei, C.~Wang, and J.~Paisley,
\newblock ``Stochastic variational inference,''
\newblock {\em The Journal of Machine Learning Research}, vol.14, no.1,
  pp.1303--1347, 2013.

\bibitem{jordan1999introduction}
M.~I. Jordan, Z.~Ghahramani, T.~S. Jaakkola, and L.~K. Saul,
\newblock ``An introduction to variational methods for graphical models,''
\newblock {\em Machine learning}, vol.37, no.2, pp.183--233, 1999.

\bibitem{blei2010bayesian}
D.~M. Blei, P.~R. Cook, and M.~Hoffman,
\newblock ``Bayesian nonparametric matrix factorization for recorded music,''
\newblock in {\em Proceedings of the 27th International Conference on Machine
  Learning (ICML-10)}, 2010, pp. 439--446.

\bibitem{lafferty2006correlated}
J.~D. Lafferty and D.~M. Blei,
\newblock ``Correlated topic models,''
\newblock in {\em Advances in neural information processing systems}, 2006, pp.
  147--154.

\bibitem{paisley2009nonparametric}
J.~Paisley and L.~Carin,
\newblock ``Nonparametric factor analysis with beta process priors,''
\newblock in {\em ACM Proceedings of the 26th Annual International Conference
  on Machine Learning}, 2009, pp. 777--784.

\bibitem{SiSEC17}
A.~Liutkus, F.~St{\"o}ter, Z.~Rafii, D.~Kitamura, B.~Rivet, N.~Ito, N.~Ono, and
  J.~Fontecave,
\newblock ``The 2016 signal separation evaluation campaign,''
\newblock in {\em Latent Variable Analysis and Signal Separation: 13th
  International Conference, LVA/ICA, Grenoble, France}, 2017, pp. 323--332.

\bibitem{nakamura2000acoustical}
S.~Nakamura, K.~Hiyane, F.~Asano, T.~Nishiura, and T.~Yamada,
\newblock ``Acoustical sound database in real environments for sound scene
  understanding and {Hands-Free} speech recognition.,''
\newblock in {\em LREC}, 2000,
\newblock [Online; accessed 29-Oct-2018] Available:
  {http://www.openslr.org/13/}.

\bibitem{kitamura2017experimental}
D.~Kitamura, N.~Ono, and H.~Saruwatari,
\newblock ``Experimental analysis of optimal window length for independent
  {low-rank} matrix analysis,''
\newblock in {\em 25th European Signal Processing Conference (EUSIPCO)}, 2017,
  pp. 1170--1174.

\bibitem{murata2001approach}
N.~Murata, S.~Ikeda, and A.~Ziehe,
\newblock ``An approach to blind source separation based on temporal structure
  of speech signals,''
\newblock {\em Neurocomputing}, vol.41, no.1-4, pp.1--24, 2001.

\bibitem{vincent2006performance}
E.~Vincent, R.~Gribonval, and C.~F{\'e}votte,
\newblock ``Performance measurement in blind audio source separation,''
\newblock {\em IEEE transactions on audio, speech, and language processing},
  vol.14, no.4, pp.1462--1469, 2006.

\end{thebibliography}
	
\end{document}